\documentclass[aps,twocolumn,prl,showpacs,amssymb,floatfix]{revtex4}
\usepackage{epsfig}

\begin{document}

\title{Instabilities in droplets spreading on gels}
\author{Karen E. Daniels$^{1,2}$, Shomeek Mukhopadhyay$^1$, Paul J. Houseworth$^2$,Robert P. Behringer$^1$}
\affiliation{$^1$Department of Physics and Center for Nonlinear and Complex Systems, Duke University, Durham, NC 27708; $^2$Department of Physics, North Carolina State University, Raleigh, NC, 27695}
\date{\today}

\begin{abstract}
We report a novel surface-tension driven instability observed for
droplets spreading on a compliant substrate.  When a droplet is
released on the surface of an agar gel, it forms arms/cracks when the
ratio of surface tension gradient to gel strength is sufficiently
large. We explore a range of gel strengths and droplet surface
tensions and find that the onset of the instability and the number of
arms depend on the ratio of surface tension to gel strength. However,
the arm length grows with an apparently universal law $L \propto
t^{3/4}$.
\end{abstract}


\maketitle

The surface-tension driven spreading of liquids is industrially
and biologically important, and has been studied in detail on both
solid and liquid substrates
\cite{Troian-1989-FIT,Cachile-1999-SSS2,Hoult-1972-OSS}.  Less is
known about how droplet spreading is modified in the presence of a
compliant substrate, a situation especially relevant to biological
applications \cite{Khanvilkar-2001-DTT,Halpern-1998-TSS}. We perform
droplet-spreading experiments on gel agar, a viscoelastic material, to
explore the influence of substrate on the spreading dynamics of the
droplet.  We find a novel branching instability with an onset that is
controlled by the ratio of surface tension difference to the shear
strength of the gel. The existence of a spreading morphology in which
a spreading droplet becomes spatially localized has important
implications for the industrial and medical application of
surfactants.  

Droplets spread differently on liquids, which are mobile, than on
solids, which are essentially rigid
\cite{Hoult-1972-OSS,Lopez-1976-SKL,Tanner-1979-SSO}. The present
experiments on spreading on a viscoelastic substrate (gel agar) are
intended as a way to span these two limits: by increasing the agar
concentration, one can tune the substrate from liquid-like to
solid-like behavior. The difference in surface tension between the
droplet (PDMS, a silicone oil; or Triton X-305, a surfactant solution)
and the substrate drives the spreading process.

\begin{figure}[bt]
\centerline{\epsfig{file=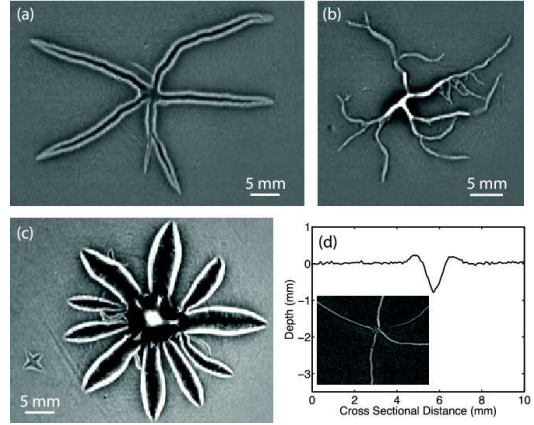, width=0.8\linewidth}}
\caption{Sample images of droplets spreading on gel agar
substrates.   (a-c) Shadowgraph images adapted from 
\cite{Daniels-2005-SWD}; bright lines represent the outer 
boundaries of the arms.
(a) Triton X-305 solution starburst: droplet concentration
$\phi=50$ ppm ($\sigma = 63$ mN/m), $X=0.12$\%w ($G = 16$ Pa), 
and $V=5\mu$L after 14 sec. 
(b) Triton X-305 solution wispy drop: droplet
concentration $\phi=100$ ppm ($\sigma = 60$ mN/m), $X=0.14$\%w ($G = 28$ Pa), 
and $V=5\mu$L after 27 sec. 
(c) PDMS starburst: $\sigma = 20$ mN/m, $X=0.08$\%w ($G = 1.3$ Pa), 
$\nu = 1000$ cS and $V=5\mu$L after 5 sec. 
(d) Scattering image of Triton X-305 starburst (inset) and 
cross section of arm.
\label{f:pics} }
\end{figure}

Previous work on the spreading of droplets on gel substrates
\cite{Szabo-2000-SLG,Kaneko-2005-KFS} showed circular spreading with
rates intermediate between those observed on solid and liquid
substrates, contrary to what we find. In addition, prior studies of
viscoelastic substrates have focused on substrate breakup or fracture
when subjected to stresses
\cite{Levermann-2002-BGP,Dufresne-2003-FFD,Vella-2005-SDF,Bonn-1998-VFR}.  
Here, we observe a
failure of the gel driven by surface tension.  After the initial
failure, there are two morphologically different manifestations of the
instability, influenced by both the substrate elasticity and the
surface tensions, which we call starbursts and wispy drops. For weak
gels (shear modulus $G \lesssim 30$ Pa), the drop breaks into distinct
crack-like spreading arms in a starburst formation, as shown in
Fig.~\ref{f:pics}a.  Morphologically, this is similar to cracking
patterns observed in \cite{Levermann-2002-BGP,Vella-2005-SDF}.  We
observe that the arms have steep sides extending into the gel
and are filled with material from the spreading droplet. Above the onset
of the starburst instability, the rate of spreading is found to be
controlled only by the width of the arms, with collapse in the data
across substrate modulus, surface tension, viscosity, and droplet
volume.  Well above the gelation transition ($G \gtrsim 30$ Pa), the
drops are typically circular as on a pre-wet solid substrate, {\itshape
i.e.} there is no indication of failure.  However, after long times
some droplets (on gels near $G \approx 30$ Pa) demonstrate the wispy
morphology shown in Fig.~\ref{f:pics}b.  We investigate the morphology
and dynamics of the starbursts in terms of droplet surface tension,
viscosity, and volume, and the strength of the gel.

The substrate is made from agar, a polymer composed of subunits of
galactose (a sugar); agar undergoes a physical gel transition for
weight concentrations of agar above $0.014$\% weight in water 
at 20.0$^\circ$C \cite{Tokita-1987-MSS}. 
We express agar concentrations as $X = M_{\mathrm
agar}/(M_{\mathrm agar} + M_{\mathrm water})$ \%w, where $M$ is the mass
of each component of the gel.  We examine droplets spreading on
$\approx 5$ mm thick gels with concentrations 0.06 $\leq X \leq$
0.14.  These values correspond to a shear modulus of 1 Pa $ \lesssim
G \lesssim$ 30 Pa \cite{Tokita-1987-MSS} and a surface tension 67 mN/m
$\lesssim \sigma \lesssim$ 75 mN/m.  We measure the latter using the
pendant drop method \cite{Andreas-1938-BTP}. For $X \lesssim 0.06$\%w the
droplets spread out as on water, and for $X \gtrsim$ 0.014\%w they remain
circular and small (mm radius) \cite{Daniels-2005-SWD}.

We use two types of droplets, Triton X-305 (octylphenoxy polyethoxy
ethanol from Dow Chemical) at concentrations $\phi$ = 1 to 1000 ppm in
deionized water and pure polydimethylsiloxane (PDMS)
at several viscosities and surface tensions. Triton X-305 is a
non-ionic surfactant with molecular weight $\approx 1526$; its
critical micellar concentration is 2000 ppm at 23$^\circ$C
\cite{Zhang-2005-AAI}. The kinematic viscosity $\nu$ of the Triton
X-305 solution is approximately that of pure water, 0.89 centistokes,
and its surface tension $\sigma_d$ ranges from that of distilled water
(73 mN/m) at the lowest $\phi$ down to 50 mN/m at $\phi=1000$
\cite{Zhang-2005-AAI}. PDMS has a density $\rho=0.968$ g/cm$^3$ and
the samples used had surface tensions $\sigma_d = 2.25$ to $20.0$ mN/m
and kinematic viscosities $\nu = 10$ to $1000$ centistokes.

We visualize the spreading droplets using shadowgraphy
\cite{Settles-2001-SST} through gel substrates prepared in clear
plastic petri dishes. A parallel beam of light passes through the
material from below; an image is formed on a ground glass above the
layer and captured by a digital camera and framegrabber.  In
addition, to visualize the arm topography we extract a calibrated
cross section using light scattering in gels containing 1 $\mu$m
polystyrene spheres \cite{Wright-1996-DLP}.  Each droplet is ejected
from a micropipette held about 1 cm above the surface of the gel. We
examined droplets with volume $V = 0.1$ to 10 $\mu$L. For reference, a
$V = 5$ $\mu$L of fluid with a molecular size of $10^{-8}$ m (PDMS),
corresponds to a monolayer surface film with radius $\approx 12$ cm.

\begin{figure}
\epsfig{file=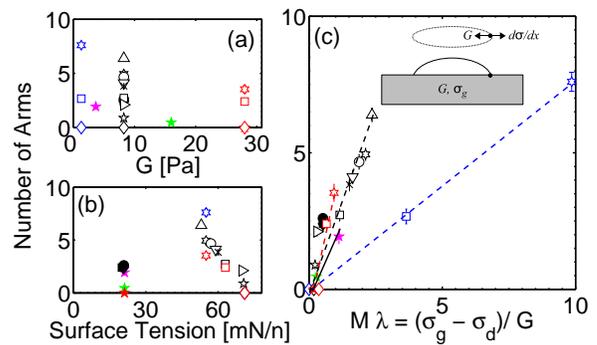,width=0.9\linewidth}
\caption{Mean number of arms on individual droplets as a function of 
(a) droplet surface tension $\sigma_d$; (b) gel elastic modulus $G$; and (c) 
$M\lambda$ (see Eq.~\ref{e:Marangoni}), assuming $\lambda_{PDMS} = 12
\lambda_{Triton}$. Inset: Schematic of Marangoni/elastic
 instability. Open symbols are Triton X-305 solutions, with
$\sigma_d$ = 72 mN/m ($\vartriangleleft$), 71 mN/m ($\diamond$), 70.6
mN/m ($\star$), 70 mN/m ($\vartriangleright$), 63 mN/m ($\square$), 60
mN/m ($\times$), 59 mN/m ($\bigtriangledown$), 57 mN/m ($\circ$), 55
mN/m ($\ast$), 53 mN/m ($\triangle$), 50 mN/m ($+$).  Solid symbols
(here overlapping) are PDMS: 
$\sigma_d = 20.1$ mN/m and $\nu = 10$ cS ($\blacksquare$),
$\sigma_d = 20.8$ mN/m and $\nu = 50$ cS ($\blacklozenge$), 
$\sigma_d = 20.9$ mN/m and $\nu = 100$ cS ($\bullet$),
$\sigma_d = 21.2$ mN/m and $\nu = 1000$ cS ($\bigstar$).
Symbol colors represent gel substrate: 1.3 Pa (blue), 3.7 Pa
 (magenta), 5.6 Pa (cyan), 8.2 Pa (black), 11.7 Pa (yellow), 16 Pa
 (green), 28 Pa (red).  Droplet volume is 2.5 $\mu$L. Each point is
 averaged over at least ten drops at the same parameters.  Horizontal
 error bars are due to $G$ and $\sigma_g$; vertical bars are the
 standard error for the measurements.  No data at $\phi=1000$ ppm is
 included, due to the unreliable counting of numerous arms. 
\label{f:onsetarms}}
\end{figure}

For the starbursts, the key physics involves a competition between
spreading due to surface tension and the  gel rigidity, 
ultimately resulting in the fracture of the gel substrate.  A droplet
on a gel surface will spread if it has a spreading parameter $\sigma_g
- \sigma_d - \sigma_{gd} > 0 $, for gel-air, droplet-air, and
gel-droplet surface tensions respectively. Since both the gel and the
droplet are largely composed of water, we assume that $\sigma_{gd}$ is
negligible  and define $\Delta \sigma \equiv \sigma_g -
\sigma_d$. The forces that drive the radial spreading of the
drop deform the gel substrate, as shown schematically in the inset to
Fig.~\ref{f:onsetarms}, and we estimate the energies involved.
Suppose that there is a nascent crack of width $d$ (corresponding
roughly to the length of a stretched polymer at the point of failure)
in the radial direction, and depth $h$ into the gel.  This crack
extends in the plane of the gel over some fraction of the outer
circumference of the drop, $aR$.  There are now several relevant
energies: the energy associated with the surface tension stretching
$E_{\sigma} \sim \Delta \sigma a R d$, the elastic energy at fracture,
$E_e \sim G a R d h$, and the thermal energy, $E_T = k_B T$.  Here, we
assume that the failure modulus of the gel is proportional to the elastic
modulus, $G$. The time to nucleate a crack, $\tau_{nuc} \sim
\exp(-E_b/E_T)$, involves an energy barrier depending on $E_e$,
$E_{\sigma}$, and $E_T$. Assuming that $d$ is comparable in size to an
extended gel polymer strand, and that $h$ is also microscopically
small, the first two energies are both many orders of magnitude larger
than $E_T$.  Consequently, we expect (and observe) rapid nucleation.
During $\tau_{nuc}$, fluid from the drop diffuses a distance $\lambda
\sim (D \tau_{nuc})^{1/2}$ into the gel.  $\lambda$ then sets the
depth for the crack: $h = \lambda$.  Therefore, we might expect a
Marangoni/elastic instability above a critical value of
\begin{equation}
M \equiv \frac{\Delta \sigma}{\lambda G} = \frac{E_{\sigma}}{E_e}.
\label{e:Marangoni}
\end{equation}

We characterize the { Marangoni/elastic} instability in terms of
several simple measures: the number of arms that form, their mean
width and their length as a function of time.  Fig.~\ref{f:onsetarms}
indicates that the parameter, $M$ captures a key feature of the
experiments, namely the number of arms.  This number increases with
$\Delta \sigma/G$; here we omit the factor $\lambda$, although we
expect that  it depends on the diffusivity of the droplet in the
gel.  However, data for such diffusivities is not available. The
number of arms increases for decreasing $G$ (connect identical
symbols, best seen for the solid stars and solid line), or for
decreasing $\sigma_d$ (for example, dashed lines connecting open
symbols of the same color). In the former case, onset occurs by
changing the gel concentration for a constant PDMS droplet, and in the
later by changing the surfactant concentration of the Triton X-305
droplet for a constant gel. There is scatter in the plot due to a
number of uncontrolled factors, for instance heterogeneities in the
gel and the impact velocity of the droplet.  Remarkably, in spite of
the three orders of magnitude difference in viscosity for the two
types of droplets, the only adjustment required to achieve coincidence
with the Triton X-305 data was to assume $\lambda_{PDMS} = 12
\lambda_{Triton}$.  This suggests that PDMS diffuses more slowly
than Triton X-305, in accordance with its longer molecular length.
An alternative framework in which to view the instability is to
consider that the number of arms in the  Marangoni/elastic
instability decreases with increased surface tension of the
droplet. This is to be expected since the wavelength of the
instability increases with surface tension in general for interfacial
phenomena \cite{Mullins-1964-SPI}.

Several limits of Eq.~\ref{e:Marangoni} are instructive. For droplets
with sufficiently high surface tension (low concentration of Triton X-305, low $\Delta \sigma$),
the central droplet does not spread significantly and the gel
remains intact, resulting in no arm growth. Similarly no arms appear
for gels with $G \gtrsim 70$ Pa; instead, the droplets spread as on a solid
\cite{Cachile-1999-SSS2,Cachile-2003-CLI}. For 30 Pa $\lesssim G \lesssim$ 70 Pa, 
a  different morphology involving the growth of thin, branching wisps was 
observed for Triton X-305 droplets (but not PDMS), as shown in
Fig.~\ref{f:pics}b and \cite{Daniels-2005-SWD}. Conversely, for gels
with $G \lesssim 1$, the substrate is essentially a viscous fluid and
the droplet spreads out rapidly, as on a deep liquid layer. In
Fig.~\ref{f:onsetarms}, note that the open blue symbols (performed on gels with
$G=1.3$ Pa) have a consistently low number of arms, suggesting that
the mechanism is nearing its limit of applicability: the gel no
longer has a yield strength. For the remainder of this work, we will focus
on the starbursts.

\begin{figure}
\epsfig{file=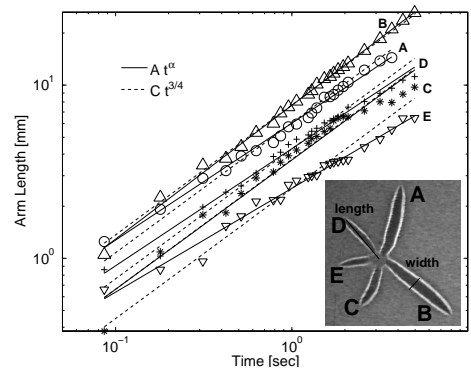,width=0.7\linewidth}
\caption{Length of the five arms (A,B,C,D,E) of a Triton X-305
starburst (inset) at a function of time. Time $t$ is measured from the
start of each arm's growth, not the drop release time. Solid lines are
fits with a free exponent; dashed lines are fixed $\alpha=3/4$. Gel is
3.7 Pa and droplet has $V= 5\mu$L and $\sigma_d = 60$ mN/m. \label{f:spreading}}
\end{figure}

\begin{figure}
\epsfig{file=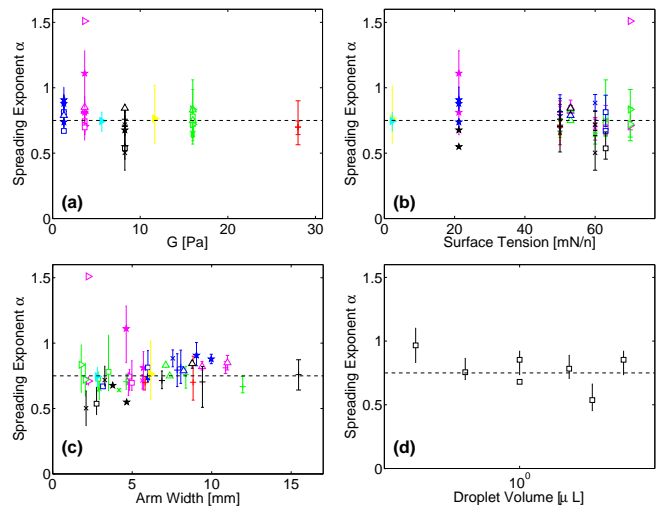, width=\linewidth}
\caption{Fit value of spreading exponent $\alpha$ vs. (a) shear
modulus $G$, for constant $V= 5\mu$L, (b) surface tension $\sigma_d$, 
for constant $V= 5\mu$L, (c) arm width, for constant $V= 5\mu$L, and (d)
droplet volume, $V$, for constant gel and droplet properties.
Points are averaged over the arms of a single
droplet. Legend as given in Fig.~\ref{f:onsetarms}. \label{f:expon}}
\end{figure}

For each of the major arms in a starburst pattern, we track the length $L(t)$ as a
function of the time from its initiation, as shown in
Fig.~\ref{f:spreading}. After $\sim 10$ seconds, the evolution of the
pattern is substantially accomplished, and the scaling discussed below
typically breaks down. For each arm, we fit $L(t)$ to $L = A
t^{\alpha}$.  Fig.~\ref{f:expon} shows that the fit value of the
exponent $\alpha$ is consistently close to $3/4$, independent of
substrate gel modulus, droplet surface tension, arm width, and
volume. In addition, the factor of 1000 in viscosity does not appear
to affect $\alpha$, which is indistinguishable for the PDMS and Triton
X-305 data.

This time dependence is the same as for oil spreading axisymmetrically
on water \cite{Hoult-1972-OSS}, where the value of the exponent arises
due to the competing effects of viscous drag and surface tension
forces. In the
liquid-on-liquid case, the prefactor
\begin{equation}
C \propto (\Delta \sigma)^{1/2} \left( \rho \mu \right)^{-1/4} 
\label{e:oilwater}
\end{equation}
has a weak dependence on the viscosity $\mu$ (as well as density
$\rho$, which we did not vary significantly in our experiments).
Intriguingly, recent work on the surfactant-induced fracture of a particle
raft \cite{Vella-2005-SDF}, saw a multi-armed morphology similar to the starbursts
observed here and $t^{3/4}$ spreading behavior, which the authors associated 
with liquid-on-liquid behavior.

\begin{figure}
\epsfig{file=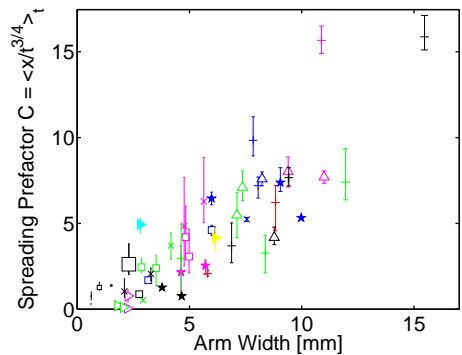, width=0.7\linewidth} 
\caption{Mean spreading prefactor $C$ as a function of mean arm
width. Each point is averaged over the arms of a single drop, with
max/min values shown as vertical bars. See legend in
Fig.~\ref{f:onsetarms} for parameters of each point. Size of symbol
indicates droplet size, with the majority taken at $V = 5\mu$L. 
\label{f:width}}
\end{figure}

In order to obtain more precise values of $C$, we assume that $\alpha
= 3/4$ is universally applicable. We then determine the time-averaged value of
the prefactor using $C = \langle L(t)/t^{3/4} \rangle_t$; this
parameter sets the scale of the spreading rate. Its value is
approximately the same for both PDMS and Triton X-305 droplets in
spite of the difference in viscosity, echoing
Eq.~\ref{e:oilwater}. Although there does not appear to be any
significant dependence of $C$ on viscosity, the average value of $C$
is linearly related to the average arm width for a given droplet.  We
note that the arm width at half-length increases and/or decreases
somewhat during arm growth, again likely due to heterogeneities in the
gel. Since the variation in width over the course of the growth is
small compared to the length of the arms, we consider its
time-averaged value. As shown in Fig.~\ref{f:width}, the wider the
arms of a droplet, the faster they grow (larger $C$).  Remarkably,
data for the various gel moduli, droplet volumes, surface tensions,
and droplet material (PDMS vs. Triton X-305) are all consistent with
this single linear relationship.

In addition to instabilities due to tangential stresses described in 
Eq.~\ref{e:Marangoni}, normal stresses due to gravity may be relevant 
as well. We consider the ratio
\begin{equation}
\frac{ \rho V g}{(\Delta \sigma) \, R}
\label{e:gravity}
\end{equation}
and find a value of about $5$, indicating that the gravitational and
surface tension effects are similar in magnitude. Although the number 
of starburst arms slowly increased with droplet volume, we observed 
starbursts for the smallest volumes that we could generate (see small 
starburst at left side of Fig.~\ref{f:pics}c).

A key issue is the nature of the mechanism through which the arms
advance. The initially smooth gel cracks under the tangential stress
from the surface tension gradient imposed by the droplet.  Once this
process has begun, the crack opens further due to both the curvature
at the tip and surface tension forces due to that curvature. The tip
velocity is set by the viscous spreading of the droplet fluid into the
opening crack, with resulting dynamics similar to that of the
liquid-on-liquid case (Eq.~\ref{e:oilwater}). Time-resolved
measurements of the initial stages of starburst formation will be
necessary to further clarify these issues. The novelty of this
cracking instability is that it is driven by an advancing layer of
fluid.



This work has been supported by NSF grant DMS-0244498 and by North
Carolina State University. We thank the UCLA/Duke/NCSU Thin Films
group,{ \it Bruno Andreotti}, and L. Mahadevan for helpful conversations.


\end{document}